\documentclass[aps,prb,twocolumn,showpacs,floatfix,superscriptaddress]{revtex4-1}
\usepackage{graphicx}
\usepackage{color}
\usepackage{amsmath}
\usepackage{amssymb}
\usepackage{amsfonts}
\usepackage{comment}
\usepackage{bm}
\usepackage{braket}
\definecolor{scarred}{rgb}{0.75,0.0,0.0}
\usepackage{hyperref}
\hypersetup{
colorlinks=true,final=true,
        linkcolor=blue,
        citecolor=blue,
        filecolor=blue,
        urlcolor=blue,
}

\begin{document}
\title{Revealing Hund's multiplets in Mott insulators under strong electric fields} 
\author{Nagamalleswararao Dasari}\email{nagamalleswararao.d@gmail.com}
\affiliation{Department of Physics, University of Erlangen-Nuremberg, 91058 Erlangen, Germany}
\author{Jiajun Li}
\affiliation{Department of Physics, University of Erlangen-Nuremberg, 91058 Erlangen, Germany}
\author{Philipp Werner}
\affiliation{Department of Physics, University of Fribourg, 1700 Fribourg, Switzerland}
\author{Martin Eckstein}\email{martin.eckstein@fau.de}
\affiliation{Department of Physics, University of Erlangen-Nuremberg, 91058 Erlangen, Germany}

\begin{abstract}

We investigate the strong-field dynamics of a paramagnetic two-band Mott insulator using real-time dynamical mean-field theory. A dielectric breakdown occurs due to many-body Landau-Zener tunnelling, with a threshold field determined by the gap. For a large range of fields, however, we predict that the tunnelling currents are small enough to allow the observation of field-induced localization of electrons, which becomes most strikingly evident in atomic-like local spin multiplets  determined by the Hund's coupling $J$.   This field-induced localization might provide a way of measuring the value of $J$ in correlated materials. It should be observable in transition metal oxides using time-resolved photo-emission spectroscopy or optical measurements in the presence of strong THz field transients. 
\end{abstract}

\maketitle

{\bf Introduction:} 
The interplay of orbital degeneracy and strong interactions is the origin of some of the most spectacular phenomena in correlated materials. An important parameter in this context is the Hund's coupling $J$, which distinguishes atomic multiplets  of different spin. The embedding of these multiplets in the solid gives rise to a plethora of interesting phenomena, including superconductivity or magnetism driven by spin or orbital fluctuations, strange metallic behavior, or metal-insulator transitions.\cite{Georges2013} However, accurate estimates of the value of $J$ in microscopic low-energy models for correlated materials are hard to obtain from first principles (using, e.g., the constrained random phase approximation\cite{Honerkamp2018,Aryasetiawan2004}), while in spectroscopic studies of solids, the local multiplet structure can be concealed. 

Ultrashort and intense laser pulses have opened intriguing new avenues to probe complex materials.\cite{Basov2017,Giannetti2016} It is therefore an obvious question whether such non-equilibrium probes can provide different means to uncover the  local interactions in the low-energy manifold of the solid. Earlier work has already shown the indirect role of multiplet excitations, or Hund's excitons, in the ultra-fast relaxation dynamics following a laser pulse.\cite{Strand2017,Rincon2018} A more direct pathway may be to explore the dynamics of electrons {\em during} an intense laser pulse. In semiconductors, atomically strong electric fields can induce coherent electron dynamics such as Zener tunnelling, high-harmonic generation, or Bloch oscillations,\cite{Schubert2014,Higuchi2017} while causing little damage to the material on the femtosecond timescale. Similar phenomena exist for Mott insulators, including many-body Landau Zener tunnelling\cite{Oka2010,Eckstein2010b,Mayer2015} the field-control of spins,\cite{Kampfrath2010} or the proposal of a distinct high-harmonic spectrum,\cite{Murakami2018,Silva2018} which exists 
despite the incoherent nature of the single-particle motion. 

A well-known quantum phenomenon related to strong fields is the Wannier-Stark localization. Independent electrons can get exponentially localized in a potential gradient, so that the energy band turns into levels separated by the potential difference between lattice sites. In metals, the electron-electron scattering can wash out this so-called Wannier-Stark ladder,\cite{Eckstein2011,Mandt2014} but in the Mott regime the localization of electrons is more  robust, until too large fields lead to a dielectric breakdown.  In the present work, we therefore first investigate the dielectric breakdown current in an orbitally degenerate Mott insulator. We then propose to reveal the local interactions in such systems using a transient field-induced localization, which may be achieved using intense THz pulses with a frequency far below the linear absorption edge. 

{\bf Model and method:} We study an orbitally degenerate Mott insulator using the two-orbital Hubbard model
\begin{align}
H= -\sum_{i,j,ll'\sigma} e^{i\phi_{ij}(t)} c^{\dagger}_{il\sigma} {\hat{T}}^{ij}_{ll'} & c_{jl'\sigma} + \sum_i {H}^\text{loc}_{i},
\label{eq:latt}
\end{align}
where $c^{\dagger}_{il\sigma}$ creates an electron at the lattice site $i$ in an orbital $l\in\{1,2\}$ with  spin $\sigma\in\{\uparrow,\downarrow\}$. The ${\hat{T}}^{ij}$'s are the hopping matrices along the bonds $(ij)$, and the electric field enters in Eq.~\eqref{eq:latt} as a time-dependent Peierls phase $\phi_{ij}=e(\vec{R}_i-\vec{R}_j)\vec{A}(t)/\hbar c$, where $\vec{A}(t)=-\int^t_0 ds \vec{E}(s)$ is the vector potential. The local term contains the Kanamori interaction\cite{Kanamori1963}
\begin{align}
{H}^\text{loc}_{i} = \,& U \sum_{l} n_{il\uparrow} n_{il\downarrow} + \sum_{\sigma,\sigma',l \neq l'} (U'-J \delta_{\sigma \sigma'}) n_{il\sigma} n_{i l' \sigma'} 
\nonumber
\\ &+ J \sum_{l \neq l'} (c^{\dagger}_{il\uparrow} c^{\dagger}_{il\downarrow} c_{il'\downarrow} c_{il'\uparrow}+c^{\dagger}_{il\uparrow} c^{\dagger}_{il'\downarrow} c_{il\downarrow} c_{il'\uparrow}),
\end{align}
where $U$, and $U'=(U-2J)$ are the intra and interorbital Coulomb interactions, respectively, and $J$ is the Hund's coupling. For two electrons, the local multiplet energies $U+J$, $U-J$, and $U-3J$ are split by $2J$.

We use real-time dynamical mean-field theory (DMFT)\cite{Freericks2006,Aoki2014} to solve this model. While the results of this paper should hold qualitatively for a generic lattice structure, we employ the specific setting introduced in Ref.~\onlinecite{Li2018}, which allows for a closed form of the DMFT self-consistency relation: This involves a Bethe lattice in which each site reflects the local environment of an atom in a cubic lattice with $e_g$ orbitals $l=d_{x^{2}-y^{2}},d_{3z^{2}-r^{2}}$, and in which the electric field acts along the body diagonal. (Fields directed along a certain crystalline direction can imply a partial field-localization, or dimensional crossover.\cite{Aron2012}) The DMFT equations for the local orbital-dependent Green's function $G_{l,l'}(t,t')$, and the current $j(t)$ are identical to Ref.~\onlinecite{Li2018}, and therefore repeated only in the appendix. The largest hopping matrix element $t_{0}=1$ sets the energy scale, so that the free half-bandwidth is $W=2t_0$. Furthermore we set $\hbar=1$ (time is measured in units of $\hbar/t_0$), and $e=1$, i.e., the field is measured in units of $t_0/ea$, where $a$ is a lattice constant. Unless otherwise stated, for all results the system is initially in equilibrium at temperature $T=0.1$ (paramagnetic phase), the Hubbard interaction is $U=8$ (Mott insulator), and the total filling of the system $n_\text{tot}=1$ (quarter filling).           
\begin{figure}[tbp]
\includegraphics{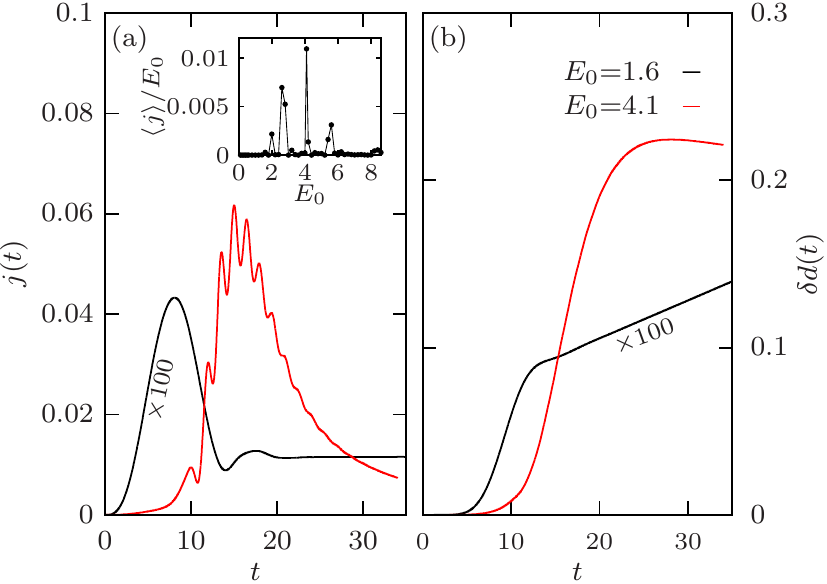}
\caption{(a) Electric current, and (b), double occupancy, for different field strengths $E_0$, at $J=0$. Inset: Current averaged over times $25 \le t \le 35$ as a function of $E_0$.}
\label{fig:tc_2}
\end{figure}
\begin{figure}[tbp]
\includegraphics{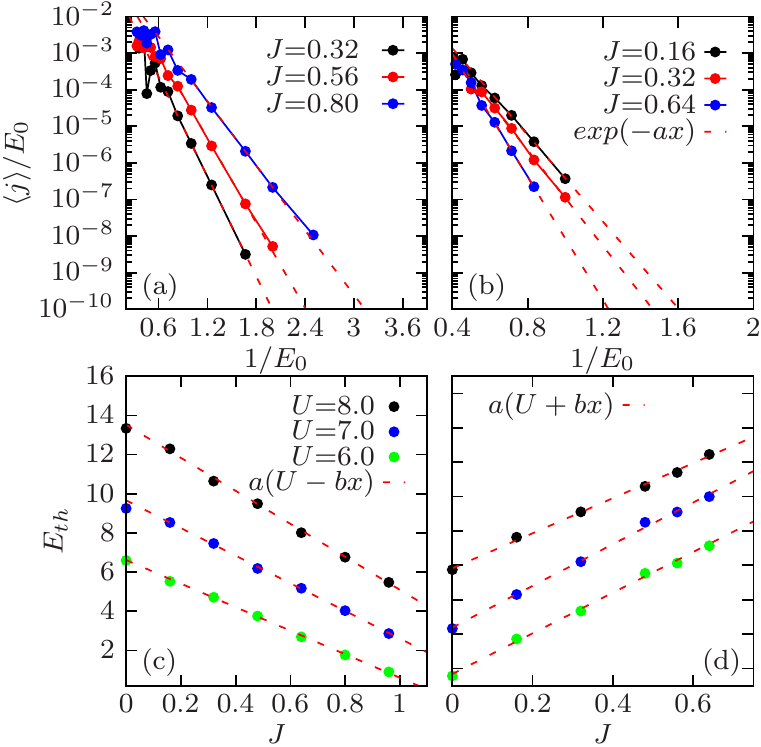}
\caption{Upper panel: Current (averaged over the times $25 < t < 35$) in the Mott insulator for two different fillings (a) $n_\text{tot}=1$ (b) $n_\text{tot}=2$. Lower panels: The threshold field $E_\text{th}$ extracted from the exponential fit for $E_0\lesssim 1$ (dashed lines in the upper panel), plotted as a function of $J$ for $n_\text{tot}=1$  (c) and  $n_\text{tot}=2$ (d).}
\label{fig:fit1}
\end{figure}

{\bf Dielectric breakdown:} To investigate the dielectric breakdown, we ramp on a field of amplitude $E_0$ within a time $t_r=15$, where $t_r$ is sufficiently longer than the inverse gap to avoid excitations. The precise field profile for $t<t_r$ is  $E(t) = E_0 [\frac12-\frac34\cos(\frac{\pi t}{t_r})+\frac14\cos(\frac{\pi t}{t_r})^3]$. The time-dependent electric current $j(t)$ and the fraction of doubly occupied sites $\delta d(t) \equiv [d(t)-d(0)] $ is plotted in Fig.~\ref{fig:tc_2}. Similar to the dielectric breakdown in the one-orbital case \cite{Eckstein2010b} we distinguish two regimes, (i) and (ii), as described below: 

(i) When $E_0$ is sufficiently smaller than the gap, the current saturates to a 
nonzero constant value after a peak during the switch-on of the field ($E_0=1.6$ in Fig.~\ref{fig:tc_2}). The initial peak corresponds to the build-up of a  polarization in the insulator, while the almost constant current at later times amounts to processes in which an electron tunnels over $\ell \propto \Delta_g/E_0$ sites in the field direction to overcome the gap energy $\Delta_g$ and create a doublon-hole pair. Correspondingly, after the switch-on, $d(t)$ increases with a rate proportional to $j/E_0$. According to this many-body Landau-Zener mechanism, the tunnelling rate and thus the current are proportional to $j/E_0\propto\exp(-E_\text{th}/{E_0})$, where the threshold field $E_\text{th}$ should scale with the gap $\Delta_g$.\cite{kirino2010,Eckstein2010b} We confirm this behavior in Fig.~\ref{fig:fit1}a) by plotting $j/E_0$ (measured at late times when the current is steady) logarithmically against $1/E_0$. Linear fits (dashed lines) indicate the exponential behavior, and  $E_\text{th}$ is extracted from the slope (see Fig.~\ref{fig:fit1}c). For given $J$, the threshold field increases with $U$,  due to an increase in the Mott gap. With increasing $J$, $E_\text{th}$ decreases linearly $E_\text{th}\approx a(U-bJ)$ as expected from 
dependence of the Mott gap on $J$.\cite{Georges2013} A rough understanding of this behavior is obtained from the multiplets in the atomic limit. Due to the  exponential dependence on the gap, the tunnelling rate is dominated by the smallest multiplet excitation $U-3J$. For a nonzero bandwidth, the  parameter $b\approx5.4,5.2,5$ for $U=6,7,8$ turns out to be different from $3$, but has the correct sign and becomes closer to $3$ with increasing $U$. For further confirmation, 
we have also performed an analogous analysis in the half-filled Hubbard model ($n_\text{tot}=2$), see Figs.~\ref{fig:fit1}b) and d). In this case, the system is initially predominantly in the high-spin doublon state (energy $U-3J$), and tunnelling creates a singly occupied and a triply occupied state with excitation energy $(3U-5J)-2(U-3J)=U+J$, which is consistent with an {\em increase} of $E_\text{th}$ with $J$ (Fig.~\ref{fig:fit1}d).

(ii) For larger fields, a strong enhancement of the field-induced excitation is observed. The doublon density quickly saturates to a large value, after which the current decays to zero, indicating the end of field-induced tunnelling ($E_0=4.1$ in Fig.~\ref{fig:tc_2}). 
Similar to the one-band Hubbard model, at $J=0$ this rapid excitation occurs at resonances $U=nE_0$ with a small integer $n$ (inset in Fig.~\ref{fig:tc_2}).  In the two-band model, once sites become doubly occupied by the field, resonances $2U=nE_0$ become visible due to the  generation of triply occupied sites from doubly occupied sites. One could therefore propose that 
the shift and splitting of these resonances for finite Hund's coupling may provide a way to measure $U$ and $J$ in the multi-band model,  but due to the broadening of the resonances and the quick heating of the system to infinite temperature, the $J$-dependence of these resonances is hard to resolve and unlikely to provide a good experimental pathway to extract $J$. For completeness, we present these data in the appendix, and  from now on focus on the spectral function as a more direct way to reveal $J$. 

\begin{figure}[tbp]
\includegraphics{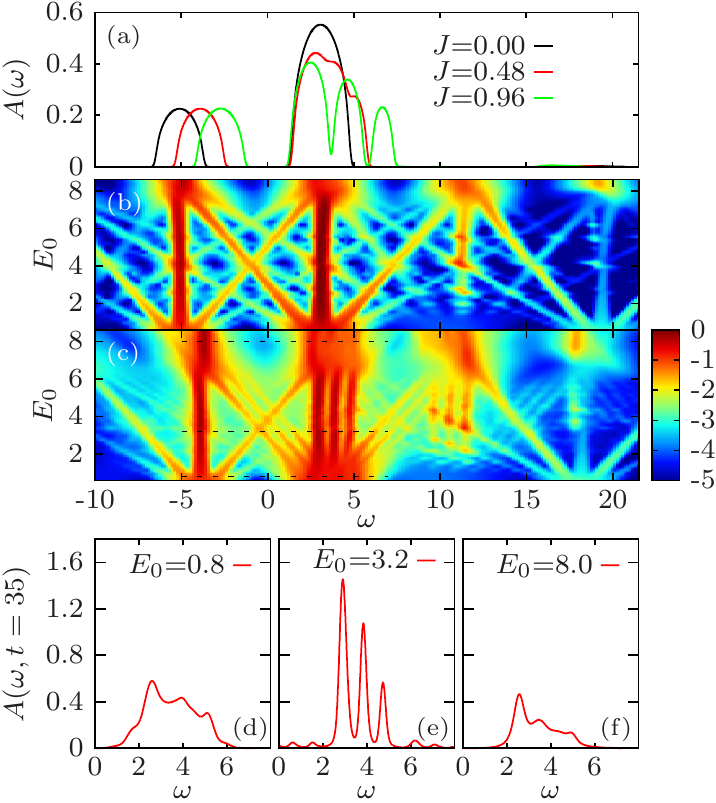}
\caption{a) Equilibrium single-particle spectral function at $U=8$ for different values of $J$. b) and c) Intensity map of the time-dependent single-particle spectral functions [$\mathrm{log}_{10}|A(\omega)|$] plotted for $J=0$ (b), and $J=0.48$ (c). Lower panel (d-f): Line plots of single-particle spectral functions plotted for the field strengths shown by dashed lines in (c).
}
\label{fig:spec}
\end{figure}

{\bf Spectral functions:}
For the same protocol as above, we have calculated the time-dependent single-particle spectral functions $A(t,\omega)=-\text{Im}{\int_0}^{s_\text{max}} ds\, G^{R}(t,t-s) e^{i \omega s}$ for different field strengths. Here $G^{R}=\sum_{l} G^{R}_{ll}$ is the orbitally averaged local propagator, and $s_\text{max}$ a cutoff set by the simulation time. In equilibrium, the spectrum has a clear Mott gap (Fig.~\ref{fig:spec}a). For very large values of $J$, the upper Hubbard band, which corresponds to transitions from predominantly singly occupied sites to the doublon multiplets, is split into three peaks separated by $2J$, but this multiplet structure is no longer visible when $J$ is sufficiently smaller than the bandwidth. When the field is turned on (Fig.~\ref{fig:spec}b for $J=0$), we observe the emergence of Wannier-stark states at energy shifts $nE_0$ from the main Hubbard bands at $\omega=\mu,U+\mu$ (and also from the weak higher-order Hubbard bands at $\omega \approx 2U-\mu,3U-\mu$), together with a band narrowing of the central peak due to the field-localization. (The band at $\omega=2U-\mu$ mainly corresponds to the insertion of an electron to a doubly occupied site, which has a substantial amplitude only once such doublons have been induced by field-induced tunnelling, while the resonance at $\omega=3U-\mu$ corresponds to the simultaneous creation of a triply occupied site and a hole upon insertion of an electron, which is possible even in equilibrium due to virtual charge fluctuations.) Weak Hund's coupling $J$ (Fig.~\ref{fig:spec}c) broadens the sidebands, and  eventually splits all peaks emerging from the Hubbard band into multiplets separated by $2J$.  This is shown clearly by three line plots of the upper Hubbard band for different electric fields, see Figs.~\ref{fig:spec}d-f). The splitting is observed for intermediate $E_0=3.2$, but disappears for large fields ($E_0 \approx U$) where the strong excitation broadens the bands. Due to the band narrowing the multiplet splitting of the Hubbard bands  is observed at much smaller values of $J$ than in the equilibrium case. 

\begin{figure}[tbp]
\includegraphics{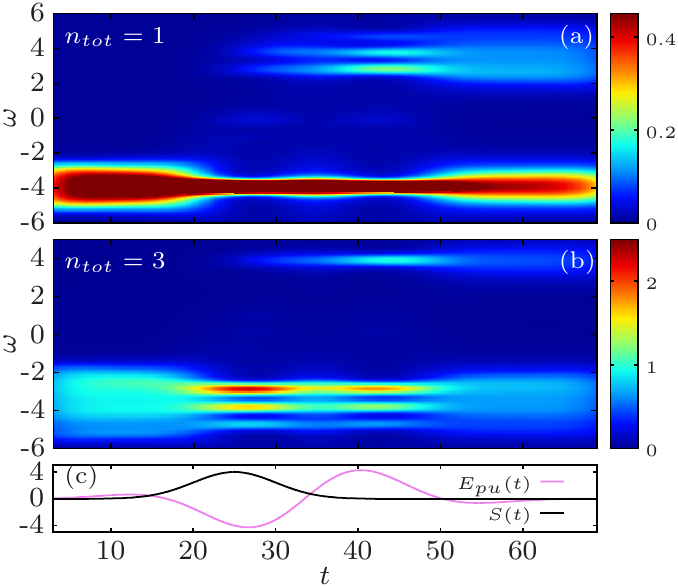}
\caption{Time-dependent photo-emission spectrum for quarter filling (a) and  three-quarter filling (b). The pump  pulse is plotted in (c), together with a probe envelope $S(t)$ of duration $t_c=5$ in arbitrary units. The interaction parameters are $U=8$, $J=0.48$, the amplitude of the pump pulse is $E_0=5.4$, and $\omega_p=0.18$.}
\label{fig:ac}
\end{figure}

The above analysis indicates that field localization can be used to measure the multiplets spectroscopically, if a strong field can be maintained for long enough without triggering a dielectric breakdown. In a realistic setup for measuring the doublon multiplets, the Mott insulator could be exposed to strong THz pulses with frequencies much smaller than the Mott gap. Here we simulate a drive of the system with a single-cycle pump pulse of the form
\begin{equation}
E_\text{pu}(t) = E_0 \sin\left[\omega_p(t-t_0)\right]e^{{-4.6(t-t_0)^2}/t_0^2},
\label{eq:pump}
\end{equation}
and then calculate the time-resolved photoemission spectrum (trPES) at different delay times $t_d$, for a Gaussian probe pulse with envelope $S(t)=\exp{(\frac{-{t}^2}{2t_{c}^2})}$ that is sufficiently shorter than the time period of the pump, using the standard expression\cite{Freericks2009} $I(\omega,t_d)\propto -i\iint dt dt' S(t-t_d)S(t'-t_d) e^{i\omega (t-t')} G^<(t,t')$. In the quarter-filled case ($n_\text{tot}=1$), one can observe a narrowing of the lower band and a splitting of the upper band,  following the time profile of the field $|E_\text{pu}(t)|$ of the pump pulse (Fig.~\ref{fig:ac}a). The upper Hubbard band corresponds to unoccupied states in equilibrium and is therefore only revealed after it has acquired some occupation due to field-induced tunnelling during the  pulse itself. In a three-quarter-filled system ($n_\text{tot}=3$, Fig.~\ref{fig:ac}b), multiplets are also seen below the Fermi energy, i.e., in the lower Hubbard band, which corresponds to transitions from mainly triply occupied sites into the doublon manifold (Fig.~\ref{fig:ac}b). 

Finally, we remark that the above expression for trPES neglects ponderomotive forces of the field on the photoelectrons, which lead to momentum shifts $\Delta k$ (and energy shifts) after the electrons have left the solid. If the probe envelope is short compared to the THz pump, these shifts are however constant for all photoelectrons emitted at the same delay $t_d$, so that the multiplet splitting remains visible and the ponderomotive effect can simply be taken into account by transforming the energy and momentum of the signal.\cite{Reimann2018} 

\begin{figure}[tbp]
\includegraphics{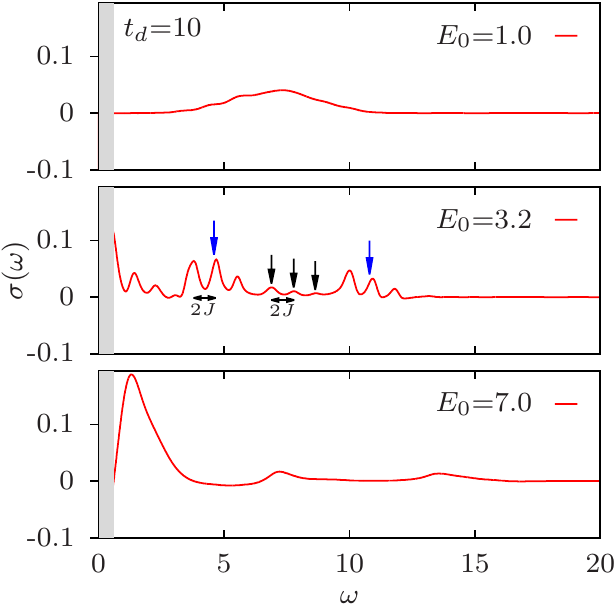}
\caption{Optical conductivity of the Mott insulator plotted for different field strengths, and a probe at delay $t_d=10$. The arrows indicate the multiplet excitations. The parameters are $U=8$ and $J=0.48$.
}
\label{fig:OC_AC}
\end{figure}

{\bf Optical response:} The multiplet splitting observed in the spectrum suggests  that similar signatures due to transitions between the various sidebands may be observable in the (experimentally more easily accessible) optical response. We have calculated the optical conductivity of the Mott insulator in a pump-probe setup using the expression\cite{Shao2016} $\sigma(\omega)=\frac{J_\text{pr}(\omega)}{E_\text{pr}(\omega)}$, where $J_\text{pr}(\omega)=J_\text{pr+pu}(\omega)-J_\text{pu}(\omega)$ is the difference in the current measured with and without the probe field $E_\text{pr}(\omega)$. The pump profile is given by the electric field used in the dielectric breakdown, and the probe is  $E_\text{pr}(t)= A_0e^{-{(t-t_d)^2}/2t_c^2}(t-t_d)/t_{c}^2$, with pulse duration $t_c$ and delay $t_d$. The probe amplitude $A_0$ is weak enough so that $\sigma(\omega)$ is measured in the limit of linear response and does not depend on $A_0$. Figure~\ref{fig:OC_AC} shows the optical conductivity of the Mott insulator for three different values of the pump field $E_0$ and time delay $t_d=10$. For small values of $E_0$, there is a broad Hubbard band around $\omega\approx7$ (Fig.~\ref{fig:OC_AC}a). A three-peak structure (shown by black arrows) with a separation of $2J$ between the peaks becomes visible for moderate field strengths (Fig.~\ref{fig:OC_AC}b). The peaks highlighted by blue arrows represent the transitions from the lower Hubbard band to Wannier-Stark sidebands of doublon multiplets. Because in the optical excitation, {\em inter}-site transitions parallel (opposite) to the field direction are dominant, the $2J$-split Wannier Stark sidebands of the Hubbard band are much stronger than the main Hubbard band in the optical conductivity. The peaks for $\omega\lesssim4$ correspond to inter Hubbard band transitions, which become possible once the system is excited. The optical conductivity around the Hubbard bands finally becomes flat for large fields ($E_0\approx U$, Fig.~\ref{fig:OC_AC}c) due to the increase of the effective temperature of the system. Even for weak fields $\sigma(\omega)$ becomes smaller at later times, due to the increas of the effective temperature (not shown). Hence, multiplet signatures in the optical conductivity can be expected in a regime of moderately strong fields, before heating starts to strongly affect the system. We have performed also simulations with transient THz pulses, like in Fig.~\ref{fig:ac}, and confirmed that multiplet signatures becomes visible for suitable parameters.

{\bf Conclusion:} 
In conclusion, we have studied a  two-band Mott insulator in strong electric field transients. The results confirm a dielectric breakdown behavior similar to the one-orbital case, with a threshold field depending on the Mott gap, and resonant enhancement of the breakdown currents around $E_0\approx (U \pm J)/n, (U-3J)/n$. 
This leaves a large parameter regime in which the Mott insulator is robust against the breakdown for times long enough so that the field-induced localization of electrons can be observed. The most striking consequence is an enhancement of the Hund's multiplets in the trPES spectrum. A related observation of how field-induced localization can reveal local interaction physics has been reported for a model of polaron formation,\cite{Werner_2015} but in the case of Hund multiplets, it should be much easier to find materials that both feature a robust gap and well-separated spectral signatures. The multiplets should also be present in the optical response, although more difficult to identify because of the various field-induced transitions which become activated in the photo-excited state. 
 While we have studied an e$_g$ system, we expect similar effects for higher orbital degeneracy, although such systems are harder to simulate numerically.  Our results suggest a pathway for exploring a distinct non-equilibrium quantum effect in the solid, which can be used to measure the Hund's coupling $J$ of a given material.  Candidates are 3d and 4d transition metal oxides with gaps and $J$ in the range of $\text{eV}$. 

This work was supported by the ERC starting grant No. 716648. The calculations have been done at the RRZE of the University Erlangen-Nuremberg.

%\bibliographystyle{apsrev4-1}
%\bibliography{apssamp}


%
% \end{document}

\newpage
\appendix*
\section{Method}

We employ the real-time Dynamical mean-field theory (DMFT) to solve the two-band Hubbard model. The DMFT formalism neglects the momentum dependence of the self-energy $\Sigma_{k}(t,t') \approx \Sigma(t,t')$, which allows to map the lattice model [Eq.~(1) in the main text] to an impurity action\cite{Freericks2006,Aoki2014}
\begin{equation*}
S=-i\int_C\!\! dt {H}^\text{loc} (t)-i\int_C dt dt' \sum_{\sigma,l,l'} {\hat{c}}^{\dagger}_{l\sigma}(t) {\hat{\Delta}}_{l,l'}(t,t') {\hat{c}}_{l'\sigma}(t')
\end{equation*}
with self-consistently determined hybridization function ${\hat{\Delta}}_{l,l'}(t,t')$. The Bethe lattice self-consistency condition under the electric field is given by\cite{Li2018}
\begin{align*}
{\hat{\Delta}}_{\sigma}(t,t') = \frac{1}{6} \!\!\!\sum_{\alpha=x,y,z,\zeta=\pm}\!\!\! e^{i \zeta \phi_{\alpha}} {\hat{T}}^{\alpha \dagger} {\hat{G}}_{\sigma} {\hat{T}}^{\alpha} e^{-i \zeta \phi_{\alpha}}
\equiv \Delta_+ + \Delta_-
\end{align*}
where quantities with a hat are  $2\times2$ matrices in orbital space. The above self-consistency condition represents a Bethe lattice in which $d$ bonds are connected to each lattice site along the three directions $\alpha= x,y,z$. The hopping matrices along the three directions are given by 
\begin{align*}
T_x= \frac{1}{4}
\begin{pmatrix} 
3 & -\sqrt{3} \\
-\sqrt{3} & 1 
\end{pmatrix},
\hspace{-0.4cm}
\quad
T_y= \frac{1}{4}
\begin{pmatrix} 
3 & \sqrt{3} \\
\sqrt{3} & 1
\end{pmatrix},
\hspace{-0.4cm}
\quad
T_z= 
\begin{pmatrix} 
0 & 0 \\
0 & 1
\end{pmatrix}.
\end{align*}
Furthermore, $\phi_\alpha$ is the Peierls phase along the bond direction $\alpha$. For  each direction there are bonds pointing along ($\zeta=+1$) and against  ($\zeta=-1$) the external field, and the corresponding contributions to  the hybridization function  are denoted by  $\hat \Delta_+$ ($\hat \Delta_-$). The multi-orbital version of the non-crossing approximation (NCA) is used to calculate the impurity self-energy $\hat \Sigma(t,t')$.\cite{Eckstein2010a} Real-time DMFT measures the local Green's function $\hat{G}$ from which the electric current can be calculated. The expression for the electric current is $j(t)$ = $\mathrm{Im} \big( \Gamma_+  - \Gamma_- \big)$ where $\Gamma_{+/-} = -i \mathrm{Tr}[\hat{\Delta}_{+/-} * \hat{G}]^{<}(t,t)$.
\begin{figure*}
\includegraphics{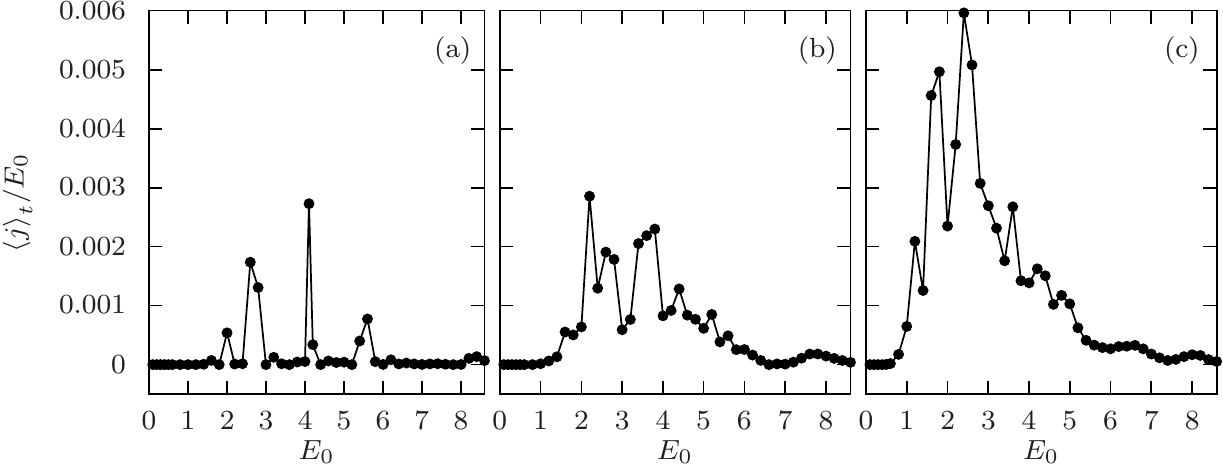}
\caption{Current averaged over time $25<t<35$, as a function of the field $E_0$ for $J=$ (a) 0.0 (b) 0.48 (c) 0.96. The interaction is $U=8$. }
\label{fig:doub_cu}
\end{figure*}

\section{Current resonances}

In this section we analyze the current and excitation density in the regime of strong fields, where near resonances $U=nE_0$ are expected to play a role. Figure~\ref{fig:doub_cu} shows the current (averaged over the time-interval $25<t<35$), for the same setting as Fig.~(1) in the main text. After an exponential increase of the current for small electric fields $E_0\lesssim1$ (i.e., the dielectric breakdown analyzed in the main text), the current becomes highly non-linear with sharp resonance peaks. For $J=0$ (Fig.~\ref{fig:doub_cu}), the strongest visible resonances appear around $E_0=U/2,U/3,U/4,2U/3$ (the resonance $E_0=U$ is not visible here, because the system is so rapidly excited that for $25<t<35$ the current has already decayed). For $J>0$, all resonances broaden and eventually merge into a continuum. The broadening is simply understood  because for $J>0$ resonances may appear at least at all combinations $E_0=(U+J)/n$, $E_0=(U-J)/n$,  $E_0=(U-3J)/n$, and are thus no longer individually resolved. 

\begin{figure*}[tbp]
\includegraphics{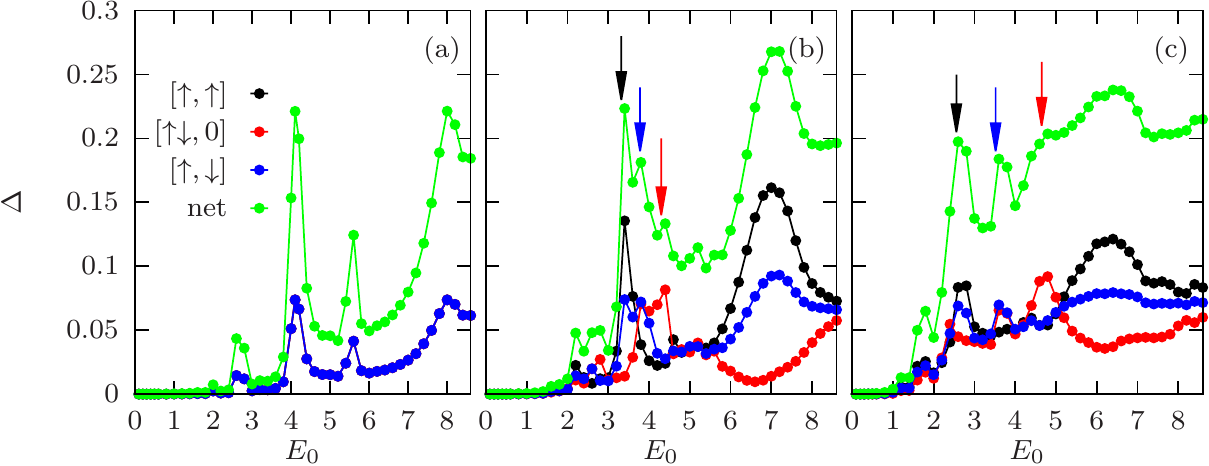}
\caption{Excitation density of doublons in different spin sectors for $J$ is (a) 0.0 (b) 0.48 (c) 0.96. Here $\Delta_{a}=d_a(t=35)-d_a(t=0)$ is the difference between excited state and equilibrium occupation of the states $a=|\!\uparrow,\uparrow\rangle$ (high-spin state),  $|\!\uparrow,\downarrow\rangle$ (inter-orbital singlet), and $|\!\!\uparrow\downarrow,0\rangle$ (intra-orbital doublon). The green curve indicates the total excitation density of doublons, and  arrows show the position of the resonances, $E_0 = (U+J)/2, (U-J)/2, (U-3J)/2$.}
\label{fig:pol}
\end{figure*}

In order to better resolve the effect of $J$ on the resonant excitation, it is slightly better to look at the total double occupancy at the impurity at a given time $t=35$ after the excitation (Fig.~\ref{fig:pol}). At least for the fields close to $E_0 = (U+J)/2, (U-J)/2, (U-3J)/2$ an enhancement of the excitation is visible (arrows in Fig.~\ref{fig:pol}b and c). Furthermore, one can resolve the individual contributions of the three multiplet occupations $|\!\uparrow,\uparrow\rangle$ (high-spin state),  $|\!\!\uparrow,\downarrow\rangle$ (inter-orbital singlet), and $|\!\!\uparrow\downarrow,0\rangle$ (intra-orbital doublon) to the double occupancy.  When $J=0$, due to the degeneracy of the atomic ground state, each spin sector contributes equally to the net excitation density of doublons. But for any finite $J$ they contribute differently to the net value, and the resonance peaks in each spin sector occur at different field strengths. For example, the single peak in Fig.~\ref{fig:pol}a) at $E_0=U/2=4$ splits into three different peaks for any finite $J$, where a maximum of $\Delta_{|\uparrow,\uparrow\rangle}$ occurs around $E_0 = (U+J)/2$, a maximum of $\Delta_{|\uparrow,\downarrow\rangle}$  occurs around $E_0 = (U-J)/2$, and a maximum of $\Delta_{|\uparrow\downarrow,0\rangle}$  occurs around $E_0 = (U-3J)/2$. 

\end{document}